\documentclass[9pt,twocolumn,twoside]{osajnl}

\journal{ol} 

\setboolean{shortarticle}{true}

\usepackage{graphicx,bm,color,mathptmx}

\newcommand{\ce}{\mathop{\mathrm{ce}} \nolimits}
\newcommand{\se}{\mathop{\mathrm{se}} \nolimits}
\newcommand{\re}{\mathop{\mathrm{Re}} \nolimits}

\newcommand{\sugg}[1]{\textcolor{red}{#1}}

\title{Fundamental quantum limits in ellipsometry}

\author[1,2]{{\L}ukasz Rudnicki}
\author[1,3,*]{Luis L. S\'anchez-Soto}
\author[1,4]{Gerd Leuchs}
\author[1,5,6]{Robert~W.~ Boyd}

\affil[1]{Max Planck Institute for the Science of Light, 
Staudtstra\ss e 2, D-91058 Erlangen, Germany}
\affil[2]{International Centre for Theory of Quantum Technologies (ICTQT), 
 University of Gda\'{n}sk, 80-308 Gda\'{n}sk, Poland}
\affil[3]{Departamento de  \'Optica, Facultad de F\'{\i}sica, 
   Universidad Complutense, E-28040~Madrid, Spain}
\affil[4]{Institute of Applied Physics, Russian Academy of Sciences, 
   603950 Nizhny Novgorod, Russia}
\affil[5]{Department of Physics, University of Ottawa, Ottawa,
   Ontario K1N 6N5, Canada} 
 \affil[6]{The Institute of Optics, University of Rochester,
  Rochester, New York 14627, USA}

\affil[*]{Corresponding author: lsanchez@fis.ucm.es}

\begin{abstract}
We establish the ultimate limits that quantum theory imposes on the accuracy attainable in optical ellipsometry. We show that the standard quantum limit, as usual reached when the incident light is in a coherent state, can be surpassed with the use of appropriate squeezed states and, for tailored beams, even pushed to the ultimate Heisenberg limit.
\end{abstract}

\setboolean{displaycopyright}{true}

\begin{document}

\maketitle

{Polatization measurements, which in a broad sense can be called polarimetry, constitute a fundamental ingredient of many optical measurement techniques~\cite{Bass:2009aa}. Polarimetry finds conceptual and practical applications in virtually every branch of science and technology.}

{Polarimetry is usually performed using a combination of wave plates and polarizers that enable direct measurements of Stokes parameters. Exhaustive research has been performed over the years on optimizing polarimetric setups~\cite{Ambirajan:1995aa,Azzam:1988aa,Sabatke:2000aa,Tyo:2002aa} and the associated sources of errors have been thoroughly identified. However, in all these analysis light is assumed to be a nonfluctuating classical field, and so the errors are exclusively related to imperfections in the setup. In other words, all of them involve technical noise  that is, in principle, subject to experimental control and can be eliminated with a proper refinement of the setup.}

{Modern schemes often involve accurate polarization measurements at faint light levels, even with single photons~\cite{Rehacek:2004aa,Ling:2006aa}. In these circumstances, quantum fluctuations of light cannot be neglected.
Actually, quantum polarimetry~\cite{Goldberg:2019aa}, as being concerned with the quantized Stokes variables, does also examine the ultimate quantum limits of their measurements.~\cite{Feng:2004aa}.} 

{In this Letter, we focus on ellipsometry, whose basis are deeply intertwined with polarimetry~\cite{Azzam:2011aa}. However, instead of Stokes parameters, the basic quantity in  ellipsometric measurements is the ellipsometric function} $\varrho$
\begin{equation}
  \label{eq:basic}
  \varrho = \frac{r_{p}}{r_{s}} = e^{i \Delta} \, \tan \psi \, ,
\end{equation}
where $r_{\sigma}$ ($\sigma \in \{ p, s \}$) are the sample's {reflection coefficients for a plane wave with the electric field polarized parallel to the plane of incidence ($p$) or perpendicular to it ($s$)}.  The parameter $\Delta$ is the {differential phase shift} between the $p$ and $s$ components upon reflection, and $\tan \psi$ is their amplitude ratio. Both, $\psi$ and $\Delta$ (and, hence, $\varrho$) can be directly determined with standard setups. {Note carefully that $\varrho$ involves only amplitude information, in contradistinction to Stokes polarimetry.} 

{Using a model-based approach, ellipsometry can determine a range of properties (including layer thickness, refractive index, morphology, and chemical composition) for films ranging in thickness from a few angstroms to several tens of microns.} These features, together with the fact that it is nondestructive, noncontact, and noninvasive, make of ellipsometry the method of choice in a variety of fields~\cite{Azzam:1987aa,Handbook:2005aa,Fujiwara:2007aa}.

For a structure of $m$ layers, the {amplitude coefficients $r_{\sigma}$ can be calculated by resorting to the transfer-matrix formalism~\cite{Sanchez-Soto:2012aa}. For a} fixed angle of incidence and wavelength, $r_{\sigma}$ depend on the material parameters ($n_i$) and layers thicknesses ($d_i$), so that one gets an involved relation $\varrho = \varrho ( n_{1}, \ldots, n_{m},d_{1}, \ldots, d_{m})$. Dispersion has to be taken into account if several wavelengths are used~\cite{Handbook:2005aa}. To infer the parameters describing the structure, this relation has to be inverted. Only a few specific cases have as yet been worked out analytically~\cite{Drolet:1994aa}. However, a vast number of numerical inversion methods have been devised which are suitable for different circumstances~\cite{Barradas:1999aa}.

Our aim {here} is to analyze how the quantum nature of light affects the precision of ellipsometric measurements. Surprisingly, these ultimate limits have not been previously examined.  In particular, we are concerned with the scaling of quantum noise with the total number of photons. We will show that settings like those based on intense coherent states are in line with the standard quantum limit~\cite{Braginskii:1975aa}, whereas an optimal phase profile of the beam given by the Mathieu function allows one to reach the Heisenberg limit~\cite{Giovannetti:2004aa}.

To introduce our model, we start by rewriting {$\rho$ as}
\begin{equation}
  \varrho  =   \frac{\mathfrak{r}_{ps}}{\mathfrak{a}_{ps}} = 
  \frac{R_{p}/R_{s}}{A_{p}/A_{s}} \, ,
  \label{eq:basic2}
\end{equation}
where $\mathfrak{r}_{ps}$ and $\mathfrak{a}_{ps}$ are the amplitude ratios, in the linear polarization basis $p$ and $s$, for the reflected ($R_{\sigma}$) and the incident ($A_{\sigma}$) fields, respectively. 

Our plan involves finding out the proper translation of \eqref{eq:basic2} into the quantum domain. This would require replacing the complex amplitudes by their appropriate quantum counterparts.  Before doing so, we observe that, since an ideal {specular} reflection only multiplies the field by a complex number, the fluctuations of the reflected field are entirely due to the fluctuations of the incident one. After all, ellipsometry, as its very name indicates, is based on the accurate determination of the polarization ellipse: if we ignore any quantum dipole fluctuations of the material system, the quantum limits are thus exclusively ruled by $\mathfrak{a}_{ps}$, which we shall consider henceforth. 

Apart from constant factors, of no relevance here, we can replace the classical amplitudes $A_{\sigma}$ with the mode annihilation operators $\hat{a}_{\sigma}$, which satisfy the bosonic commutation relations $[\hat{a}_{\sigma}, \hat{a}^{\dagger}_{\sigma^{\prime}}] = \delta_{\sigma \sigma^{\prime}}$, ($\sigma, \sigma^{\prime} \in \{p, s \}$ as before). We thus have
\begin{equation}
  \label{eq:rhoi1}
  \hat{\mathfrak{a}}_{ps} =  \frac{\hat{a}_{p}}{\hat{a}_{s}} =
  \hat{a}_{p} \hat{a}_{s}^{\dagger} 
  ( \hat{a}_{s} \hat{a}_{s}^{\dagger} )^{-1} =
  \hat{a}_{p} \hat{a}_{s}^{\dagger}  (\hat{N}_{s} +1 )^{-1} \, ,
\end{equation}
where $\hat{N}_{\sigma} = \hat{a}^{\dagger}_{\sigma} \hat{a}_{\sigma}$ are the number operators for each basic polarization mode. Please observe carefully  that the quotient $\hat{a}_{p}/\hat{a}_{s}$ is meaningful, since there is no problem with the ordering of operators. Similar amplitude ratios have been considered before to deal with quantum polarization~\cite{Singh:2013aa}.

Next, following a well-established procedure~\cite{Luis:1993aa}, we decompose the amplitudes as $\hat{a}_{p}\hat{a}_{s}^{\dagger} = \hat{E}  [\hat{N}_{p}\left(\hat{N}_{s} + 1 \right)]^{1/2}$, where $\hat{E}$ is a unitary operator that represents the exponential of the relative phase between the modes $p$ and $s$. In this way, we can recast~\eqref{eq:rhoi1} as
\begin{equation}
  \label{eq:poldec}
  \hat{\mathfrak{a}}_{ps} =  \hat{E} \; \hat{P} \, ,
  \qquad
  \hat{P} = \sqrt{\frac{\hat{N}_{p}}{\hat{N}_{s}+1}} \, .
\end{equation}
Since $\hat{P}$ is a positive-semidefinite operator, the polar decomposition \eqref{eq:poldec} can be seen as the quantum version of the factorization in \eqref{eq:basic} in terms of a phase $\hat{E}$ (which plays the role of $e^{i \Delta}$) and a modulus $\hat{P}$ (the analogous to $\tan \psi$)~\cite{Halmos:1982aa}, applied to the incident field. {Observe that the commutation relations force the appearance of $\hat{N}_{s} + 1$ instead or $\hat{N}_{s}$ in the denominator of $\hat{P}$, which breaks an apparent symmetry in the classical definition of $\varrho$ under the interchange of modes $p \leftrightarrow s$.} 

To examine the properties of $\hat{E}$ and $\hat{P}$, we introduce two new operators 
\begin{equation} 
\hat{N} = \hat{N}_{p} + \hat{N}_{s} \qquad
\hat{L} = \frac{1}{2} ( \hat{N}_{p} - \hat{N}_{s}) \, , 
\end{equation}
which correspond to the total photon number and (apart from the factor 1/2) the photon number difference between the two modes.  Since $[\hat{N}, \hat{E}]= 0$, we can study the restrictions $\hat{E}^{(N)}$ to each subspace with fixed number of photons, which have been aptly termed as Fock layers~\cite{Muller:2016aa}. If we denote the Fock basis of the two modes as $| m , n \rangle = | m \rangle_{p}
\otimes |n \rangle_{s}$, the restriction  $\hat{E}^{(N)}$ turns out to be~\cite{Luis:1993aa}
\begin{equation}\label{phaseN}
  \hat{E}^{(N)} = \sum_{n=0}^{N-1}
  | n,N-n \rangle \langle n+1,N-n-1| + | N,0 \rangle\langle 0,N| \, . 
\end{equation}
The extra contribution $| N,0 \rangle\langle 0,N| $, related to the quantum vacuum,  makes $\hat{E}^{(N)}$ unitary in the  $N$-photon layer. {The total operator $\hat{E}$ is obtained by summing over all the Fock layers $\hat{E} = \sum_{N} \hat{E}^{(N)}$; it is unitary} and defines a Hermitian relative phase via $\hat{E} = \exp (i \hat{\Phi})$. Interestingly, $\hat{\Phi}$ has a discrete spectrum: for each Fock layer, there are $N+1$ uniformly distributed eigenvalues in the interval $[0, 2 \pi]$.  When $N$ is large, this spectrum becomes dense and we can take this variable as continuous. This is the limit we shall consider in what follows, as it is the situation encountered in most of realistic ellipsometric experiments.

To elucidate this situation in more detail, it will prove convenient to relabel the Fock basis $|m, n \rangle$ in terms of the common eigenstates of $\hat{N}$ and $\hat{L}$ (note that $[\hat{N}, \hat{L} ] = 0$): $|N, \ell \rangle$, with $\ell=-N/2, \ldots, N/2$.  When $N \gg1$ this basis is effectively infinite dimensional and, to simplify the notation, we will omit $N$  and label these  states just  by $| \ell \rangle$.  The action of the unitary operator $\hat{E}$ in the $| \ell \rangle $ basis is $\hat{E} | \ell \rangle = | \ell -1 \rangle$ and, in the representation generated by the normalized eigenvectors of  $\hat{E}$,  we have  
\begin{equation}
\label{eq:canrep}
\hat{L} \mapsto -i\partial_{\phi}\, , 
\qquad 
\hat{E} \mapsto e^{i \phi} \, ,
\end{equation} 
as it happens for the canonical pair angle-angular momentum~\cite{Hradil:2006aa,Rehacek:2008ss}.

The relative-phase wave function $\Psi( \phi) = \langle \phi | \Psi \rangle$ defines a continuous probability density $ p(\phi)=|\Psi(\phi)|^2$ that is Fourier related with the basis $|\ell \rangle$; namely,
\begin{equation}
\label{eq:conphil}
  \Psi (\phi)=\frac{1}{\sqrt{2\pi}} 
  \sum_{\ell=-\infty}^{\infty}
  e^{-i  \ell\phi}  \, \Psi_{\ell} \, ,
\end{equation}
with $\Psi_{\ell}= \langle \ell | \Psi \rangle$.

In principle, {every quantum state has an expansion in the number basis and therefore spans several Fock layers (leaving aside the number states).} Since there are no coherences across them, when $N \gg 1$ we can replace the action of the operator $\hat{N}$ by its average $\bar{N}$. In addition, we take $\langle \hat{L} \rangle \ll \bar{N}$; this holds when the sample's reflectivity is high, which holds in most practical cases. We stress though that this hypothesis simplifies the calculations, but it is unessential for our results. We have now that \eqref{eq:poldec} can be rewritten as
\begin{equation}
  \hat{P}  \simeq 1 +
  \frac{2}{\bar N} \hat{L} \, ,
\end{equation}
which shows {that the relevant variable in this limit is $\hat{L}$.}

From this perspective, ellipsometry reduces to the simultaneous measurement of both $\hat{E}$ and $\hat{L}$.  If the second vacuum-related contribution  in~\eqref{phaseN} can be neglected (which happens, as we have said, for all the situations of interest), we can use the representation \eqref{eq:canrep}. In this limit, these operators satisfy the commutation relation~{$[ \hat{E} , \hat{L} ] = \hat{E}$}, which immediately leads to an uncertainty relation that reflects the fact that both magnitudes cannot be simultaneously measured with arbitrary precision.

Since $\hat{E}$ is unitary, the notion of variance must be accordingly adapted~\cite{Levy:1976aa}: {$\Delta^{2} \hat{E} = \langle \hat{E}^\dagger \hat{E} \rangle -  \langle \hat{E}^\dagger \rangle \langle \hat{E} \rangle = 
1 - | \langle \hat{E} \rangle |^{2}$}.  This coincides with the circular variance, which is the proper way of dealing with a periodic variable in statistics~\cite{Rao:1965aa}.  With this alternative standpoint, the usual form of the uncertainty relation; viz, $\Delta^{2} \hat{A} \, \Delta^{2} \hat{B} \ge | \langle [\hat{A}, \hat{B} ] \rangle |^2/4$, becomes
\begin{equation}
  \label{disp}
  \Delta^{2} \hat{E}  \; \Delta^{2} \hat{L}  \ge
  \frac{1}{4} |\langle \hat{E} \rangle|^2 \, .
\end{equation}

Before we proceed with a systematic treatment, let us examine a natural choice for the input state:  the  two-mode coherent state
$|\alpha_{p} , \alpha_{s} \rangle$. A direct calculation gives
\begin{eqnarray}
\label{Lcoh}
  &\displaystyle \Delta_{\mathrm{coh}}^{2} \hat{L}   =  
  \frac{|\alpha_{p}|^2+|\alpha_{s}|^2}{4} =   \frac{\bar N}{4} \, , & \nonumber \\
  & & \\
  & \displaystyle \langle \hat{E} \rangle_{\mathrm{coh}}  \simeq 
  e^{i(\phi_{p} - \phi_{s})} 
  \left(1 - \frac{1}{8|\alpha_{p}|^2} \right)
  \left(1 - \frac{1}{8|\alpha_{s}|^2}\right) \, , & \nonumber
  \end{eqnarray}
with  $\alpha_{\sigma}=|\alpha_{\sigma}|e^{i \phi_{\sigma}}$ and, in the second equation, we have assumed large $| \alpha_{\sigma}|$. 
In the optimal choice $|\alpha_{p}| = |\alpha_{s}|=\sqrt{\bar N/2}$, this result boils down to $\Delta_{\mathrm{coh}}^{2} \hat{E} \simeq  1/\bar{N}$ and, therefore, in the limit $\bar{N} \gg 1$, these states do saturate the uncertainty relation (\ref{disp}).  As could be anticipated, this is the standard quantum limit for ellipsometry; i.e., the uncertainty of $\varrho$ scales with $1/\sqrt{\bar{N}}$. This statement follows from the fact that while both $\langle \hat{E} \rangle_{\mathrm{coh}}$ and $\langle \hat{P} \rangle_{\mathrm{coh}}$ do not scale with $\bar{N}$, the uncertainties $\Delta_{\mathrm{coh}} \hat{E}$ and $\Delta_{\mathrm{coh}} \hat{P}$ do scale like $1/\sqrt{\bar{N}}$. Therefore, the accuracy of the relative phase, which renders the accuracy of $\hat{L}$, fully depends on the number of photons. In this way, setting the accuracy of the relative phase fixes the average  number of photons, leaving no room for improvements of the scaling property of $\Delta_{\mathrm{coh}}^{2} \hat{L}$.

The treatment of the previous paragraph assumed that the input state is separable. One might naively expect that entangling the $p$ and $s$ modes would make it possible to bypass the standard quantum limit.  In this vein,  a natural choice is a two-mode squeezed state, $|\alpha_{p} , \alpha_{s},\zeta \rangle$, which  is a displaced squeezed  vacuum with a complex squeezing parameter $\zeta= \mathfrak{s} e^{i\theta}$. Using the results for the second-order moments of the photon numbers~\cite{Lee:1990aa}, we get 
\begin{equation}
  \Delta_{\mathrm{sq}}^{2} \hat{L}  = \tfrac{1}{4} 
  \left [ (|\alpha_{p}|^2+|\alpha_{s}|^2) \, \cosh(2\mathfrak{s} ) -
  2 |\alpha_{p} \alpha_{s}| 
  \cos( \delta \phi )\, \sinh(2\mathfrak{s} ) \right] \, ,
\end{equation}
with $\delta \phi= \phi_{p}+ \phi_{s} - \theta$. In the optimal setting, when $\delta \phi=0$ and $|\alpha_{p}|=|\alpha _{s}|=\sqrt{\bar N/2-\sinh^{2}\mathfrak{s}}$, we obtain
\begin{eqnarray}
  \label{eq:noisq}
  & \displaystyle \Delta_{\mathrm{sq}}^{2} \hat{L}  = \tfrac{1}{4} 
  (\bar{N}-2\sinh^{2}\mathfrak{s}) e^{-2\mathfrak{s}} 
  \lesssim \frac{\bar{N}}{4} e^{-2\mathfrak{s}} \, , & \nonumber \\
  & & \\
& \displaystyle \langle \hat{E} \rangle_{\mathrm{sq}} \simeq e^{i(\phi_{p} - \phi_{s})} 
\left ( 1 -  \frac{2 \sinh^{2} \mathfrak{s}}{\bar{N}} \right) \, . & \nonumber
\end{eqnarray} 
where in the second equation we have utilized the approximation $\hat{E} \simeq 2 \hat{a}_{p} \hat{a}_{s}^{\dagger}/\bar{N}$, which works well for large squeezing. In this regime, we effectively get $\Delta_{\mathrm{sq}}^{2} \hat{E}=e^{2\mathfrak{s}}/\bar{N}$, confirming that the uncertainty relation (\ref{disp}) is saturated.  In interferometry, squeezed states allow us to beat the standard quantum limit by reducing the noise in one quadrature at the expense of increasing the noise in the conjugate quadrature~\cite{Caves:1981aa}. Much in the same way, \eqref{eq:noisq} shows that we can control the quantum-noise balance between $\hat{E}$ and $\hat{L}$. For an experimental scheme and a particular system under study, one can perform a conventional analysis of the sensitivity of the parameters to be estimated to the noise in the measured $\psi$ and $\Delta$. Redistributing the noise between these variables is a resource to improve the practical precision.

Actually, ellipsometric measurements are limited by shot noise, particularly at low light intensities or when using ellipsometers employing a nulling technique.  The use of entangled beams in ellipsometry has been previously reported~\cite{Abouraddy:2001aa,Toussaint:2004aa} and it was shown how this technique can improve present standards.

Let us now go back to the uncertainty relation (\ref{disp}).  States satisfying the equality in an uncertainty relation are sometimes referred to as intelligent states.  The left-hand side can be minimized (getting the value $0$) for eigenstates of $\hat{L}$. However, this situation is trivial: since the right-hand side must vanish as well, it follows that $\langle \hat{E} \rangle=0$. 

The two previous examples of coherent and squeezed states evidence that \eqref{disp} can be saturated in the limit of intense fields $\bar{N} \rightarrow \infty$. However, it is well known that for the general case of finite $N$ this bound cannot be exactly attained. Therefore, we modify our strategy and  look  instead for normalized states that minimize the uncertainty product $\Delta^{2} \hat{E} \ \Delta^{2} \hat{L}$ under the condition that $\langle \hat{E} \rangle$ and $\langle\hat{L}\rangle$ are fixed (albeit \emph{a priori} unknown) parameters. As a consequence, what is left for optimization is $\langle\hat{L}^2\rangle$.

We approach this problem by the method of undetermined multipliers. The linear combination of variations leads to the basic equation~\cite{Hradil:2006aa}
\begin{equation}
  \label{eigcom}
  [ \hat{L}^2 +  \mu \hat{L} + \tfrac{1}{4} ( q^\ast \hat{E} + q
  \hat{E}^\dagger )] | \Psi \rangle = a | \Psi \rangle ,
\end{equation} 
where $\mu$, $q$, and $a$ are Lagrange multipliers. The factor of $1/4$ was included for  convenience. We  solve this equation in the phase representation $\Psi (\phi) = \langle \phi | \Psi \rangle$. For simplicity, we also take $q$ to be real and  nonnegative, since its argument is the phase of $\langle \hat{E} \rangle$ and, as such,  can be reintroduced whenever necessary. {With the change of variable $\Psi (\phi)=e^{i \mu \eta} \tilde{\Psi}(\eta)$, with $\eta= \phi/2$, we arrive at the Mathieu equation~\cite{McLachlan:1947aa}}
\begin{equation}
  \label{Mathieu}
  \frac{d^2 \tilde\Psi (\eta)}{d\eta^2}  + [ \tilde a - 2 q
  \cos (2 \eta) ] \ \tilde\Psi (\eta) = 0,
\end{equation}
with $\tilde a= 4a+\mu^2$. The variable $\eta$ has a domain $ 0 \le \eta < 2 \pi$ and plays the role of polar angle in elliptic coordinates. In our case, the required periodicity of  $\phi$ imposes that the only acceptable Mathieu functions are those being periodic with the period of $\pi$ in $\eta$. The values of $\tilde a$ in Eq.~(\ref{Mathieu}) that satisfy this condition are the eigenvalues of
this equation. 

We have then two families of independent solutions, namely the angular Mathieu functions $\ce_{k} ( \eta, q) $ and $\se_{k+1} (\eta, q)$ with $k = 0, 1, 2, \ldots$, which are usually known as the elliptic cosine and sine, respectively. The eigenvalues associated with these solutions are conventionally denoted as $a_k(q)$ and $b_{k+1}(q)$. The parity of both eigenfunctions is exactly the same as their trigonometric counterparts, that is, the elliptic cosines are even while the elliptic sines are odd in $\eta$. Both functions have the period $\pi$ when their index ($k$ or $k+1$ respectively) is even or period $2 \pi$ when it is odd. Thus, the acceptable solutions for our problem are the independent Mathieu functions of the even order.

Because of the above symmetry properties, we can easily find that
$\langle\hat L\rangle=-\mu/2$, which further specifies the phase of
$\Psi (\phi)$ to be $e^{-i \langle\hat L\rangle \phi}$. Finally, we
obtain ($ k = 0, 1, \ldots$)
\begin{equation}
  \label{Matsol}
  \Psi_{k} (\phi, q) =
  \frac{e^{-i \langle\hat L\rangle \phi}}{\sqrt{\pi}}
  \left \{
    \begin{array}{ll}
      \ce_{2k} (\phi/2, q ) , &  \\
      \se_{2k+2} (\phi/2, q ) ,
    \end{array}
  \right . 
\end{equation}
where the factor $1/\sqrt{\pi}$ ensures proper normalization on the
interval $ 0 \le \phi < 2 \pi$.

We consider only even solutions, although a parallel treatment can be done for the odd ones. {After some calculations we obtain 
\begin{equation}
  \Delta_{k}^{2} \hat{L} =  
   \tfrac{1}{4} \left[ A_{2k}(q) - 2q \, \re  \ \Theta_{k} (q)\right]\, ,
   \qquad
  \Delta_{k}^{2} \hat{E}  =  1 - |\Theta_{2k} (q)|^{2} \, ,
  \label{eq:eqgen}
\end{equation}
where} $\Theta_{k} (q)  = A_{0}^{(2k)} (q) \; A_{2}^{(2k)} (q) +  \sum_{j=0}^{\infty} A_{2j}^{(2k)} (q) \; A_{2j+2}^{(2k)} (q)$   and the coefficients $A_{j}^{(k)} (q)$ are defined in terms of the expansion $\ce_{k} (\eta, q) = \sum_{j=0}^\infty A_{j}^{(k)} (q) \, \cos (j \eta)$, so they determine the Fourier spectrum and satisfy recurrence relations that can be efficiently computed by a variety of methods.

\begin{figure}[t]
  \centerline{\includegraphics[width=.85\columnwidth]{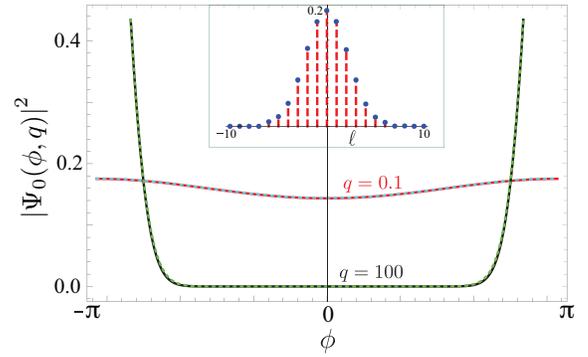}}
  \caption{Probability density of the relative phase for the fundamental Mathieu wave function $\Psi_{0} (\phi, q)$, saturating the uncertainty relation~(\ref{disp}), for two different values of the phase dispersion $q$. The continuous lines correspond to the true probability, whereas the dotted lines are the von Mises approximations as in \eqref{eq:Misesapprox}. In the inset, we show the Fourier components $|\Psi_{\ell}|^{2}$ corresponding to $q=0.1$.}
  \label{fig:comp}
\end{figure}

Formulas (\ref{eq:eqgen}) can be studied by means of both numerical methods and analytical considerations based on asymptotic expansions of the Mathieu functions. These asymptotic limits identify the fundamental mode $k=0$ as the minimum uncertainty state for all the values of the parameter $q$ and, henceforth, $\Psi_0 (\phi, q)$ is the solution we were looking for.

The corresponding probability density $p(\phi) = |\Psi_{0} (\phi, q)|^{2}$ can be approximated by 
\begin{equation}
\label{eq:Misesapprox}
p(\phi) = \frac{1}{\pi} |\ce_0 (\phi/2, q)|^2 \simeq \frac{1}{\pi} 
\left \{
\begin{array}{ll}
\exp (- q \cos \phi ),  & q \rightarrow 0 , \\
\exp (- \sqrt{q} \cos \phi ) & q \rightarrow \infty .
\end{array}
\right .
\end{equation} 
In both  limits, this $p(\phi ) $  may be approximated by a von Mises distribution, $p(\phi ) \propto \exp [ - \kappa \cos (\phi - \phi_{0}) ]$, which is considered as the circular analog of the Gaussian distribution~\cite{Mardia:2000aa}.  The parameter $\phi_{0}$ is the mean phase, while $\kappa$ (which is directly related to $q$) is a measure of concentration (i.e., a reciprocal measure of dispersion). If $\kappa$ tends to zero, the distribution is close to uniform, whereas when  $\kappa$  is large, the distribution becomes very concentrated.  We thus conclude that von Mises wave functions constitute an excellent approximation to the fundamental Mathieu wave function, except perhaps for intermediate values of the dispersions, where a deviation may occur. This behavior is illustrated in Fig.~\ref{fig:comp}, where we compare $p(\phi)$ for two extreme values of $q$. We also plot the Fourier components $\Psi_{\ell}$ of the state, defined via \eqref{eq:conphil}. In this way,  we have characterized optimal input states for which the relative phase $\phi$ between $p$ and $s$ components is  continuously distributed with {probability $p(\phi)$.}

Most importantly, the bound (\ref{disp}) can now be saturated independently of the value of $\bar{N}$. The accuracy of the phase  becomes fixed by the choice of $q$ which is the inverse of the  Gaussian width, while Eq.~(\ref{disp}) provides the uncertainty of $\hat L$ as being the function of $|\langle e^{i \phi} \rangle|$ only. At the level of $\hat{E}$ and $\hat{L}$ the situation seems to be analogous to that of the coherent and squeezed states. However, the lack of dependence on the photon number, which now is an external parameter absent in the wavefunction $\Psi_{0}$, leads to the Heisenberg scaling: $\bar{N}^{-2}$ when it comes to the uncertainty of the modulus $\Delta^{2} \hat{L}$ and consequently the $1/\bar{N}$ scaling for the uncertainty of $\varrho$.

To conclude, it is interesting to look at the optimal states discussed thus far from the perspective of polarization squeezing, which can be seen as a continuous-variable polarization entanglement.  For $\bar{N} \gg 1$ the standard Stokes operators~\cite{Luis:2000ys} can be approximated as
{$\hat{S}_{z} = \hat{L}$, $\hat{S}_{+} = \bar{N} \hat{E}^{\dagger}$,
and $\hat{S}_{-} = \bar{N} \hat{E}$}, with $\hat{S}_{\pm} = \hat{S}_{x} \pm i \hat{S}_{y}$. Polarization squeezing occurs when~\cite{Chirkin:1993dz,Muller:2012ys} $\bar{N}\Delta^{2} \hat{S}_z/(|\langle\hat{S}_x\rangle|^2+|\langle \hat{S}_y\rangle|^2)<1$, which in our context can be simply reformulated as ${\Delta^{2} \hat{L} < 
  \tfrac{1}{4} \bar{N} |\langle\hat{E}\rangle|^2}$. A glance at Eqs.~(\ref{Lcoh}) and (\ref{eq:noisq}) reveals that the coherent states are not polarization squeezed, but the squeezed states $|\alpha_p , \alpha_s,\zeta \rangle$ do present substantial amount of polarization squeezing. On the other hand, the optimal Mathieu beams, $\Psi_{0} (\phi, q)$,  are polarization squeezed whenever $\sqrt{q}<\bar{N}$.  \sugg{For them, the} average value of the Stokes vector is given by the free parameters of the state.

{The ideal squeezed states require an infinite amount of energy and they can therefore not be generated in the lab.  The squeezed states that can significantly improve the performance of a delicate measurement, such as in the case of gravitational wave detection, are always states showing finite squeezing, which nevertheless may be high. The same is true here. One purpose of this manuscript is to discuss the improvement such states offer in the case of ellipsometry. Another purpose it that in this particular application, we found that by using special states of finite energy we can do even a better than with squeezed states of the same energy and we provide their mathematical properties. The possibility of creating them in the lab is still under study.} 

In summary, we have investigated how unavoidable quantum noise limits the accuracy of ellipsometric measurements. Coherent states are shot-noise limited, whereas squeezed states achieve the Heisenberg scaling only in the limit of very large $N$. However, we have found a set of states, with a Mathieu wave function, which yield the optimal scaling precisely in the moderate-light regime. This regime has been ignored thus far by classical analysis but, as quantum technologies improve, the use of entanglement and squeezing to enhance precision in ellipsometry is likely to become more widespread.

\noindent \textbf{Funding.} Foundation for
Polish Science (ICTQT 2018/MAB/5); 
Ministry of Education and Science of the Russian Federation (14.W03.31.0032);
Canada Excellence Research Chairs (501100002781).
Ministerio de Ciencia e Innovaci\'on (PGC2018-099183-B-I00)

\noindent\textbf{Disclosures.} 
The authors declare no conflicts of interest.

\bibliography{ellipsometry}
\bibliographyfullrefs{ellipsometry}

 \newpage

\end{document}